\documentstyle[prb,preprint,aps]{revtex}
\begin{document}
\draft
\title{
Aharonov-Bohm Oscillations in a One-Dimensional Wigner Crystal-Ring
}
\author{
I.~V.~Krive$^{1,2}$, R.~I.~Shekhter$^{1}$, S.~M.~Girvin$^{3}$,
and M.~Jonson$^{1}$ }
\address{
$^{(1)}$Department of Applied Physics, Chalmers University of
Technology and G\"{o}teborg University, S-412 96 G\"{o}teborg,
Sweden }
\address{
$^{(2)}$B. I. Verkin Institute for Low Temperature Physics and Engineering,
47 Lenin Avenue, 310164 Kharkov, Ukraine.
}
\address{
$^{(3)}$Department of Physics, Indiana University, Bloomington, IN
47405}

\maketitle
\newpage
\begin{abstract}
We calculate the magnetic moment (`persistent current') in a strongly
correlated electron system --- a Wigner crystal --- in a one-dimensional
ballistic ring. The flux and
temperature dependence of the persistent current in a perfect ring is
shown to be essentially the same
as for a system of non-interacting electrons.
In contrast, by incorporating into the ring geometry
a tunnel barrier that pins the Wigner
crystal, the current is suppressed and its
temperature dependence is drastically changed. The competition between
two temperature effects --- the reduced barrier height for macroscopic
tunneling
and  loss of quantum coherence ---  may result in a
sharp peak in the temperature dependence.
The character of the macroscopic quantum tunneling of a Wigner crystal ring
is dictated by the strength of pinning. At strong pinning the tunneling
of a rigid Wigner crystal chain is highly inhomogeneous, and the persistent
current has a well-defined peak at $T\sim 0.5\ \hbar s/L$ independent of the
barrier height ($s$ is the sound velocity of the Wigner crystal, $L$ is the
length of the ring). In the weak pinning regime, the Wigner crystal tunnels
through the barrier as a whole and if $V_p>T_0$ the effect of the barrier
is to suppress the current amplitude and to shift the
crossover temperature from $T_0$ to $T^*\simeq \sqrt{V_{p}T_{0}}$.
($V_{p}$ is the amplitude of the pinning potential, $T_{0} =\hbar v_{F}/L ,\;
v_{F}\sim \hbar/ma $ is the drift velocity of a Wigner crystal ring with
lattice spacing $a$). For
very weak pinning, $V_p\ll T_0$, the influence of the barrier on the persistent
current of a Wigner crystal ring is negligibly small.
\end{abstract}

\section{Introduction}

In multiply-connected condensed matter systems, the Aharonov-Bohm effect
manifests itself as oscillations of both thermodynamic and kinetic quantities
with respect to the  enclosed
magnetic flux, $\Phi$ (for a review see Ref.~\onlinecite{Washburn}).
{}From a purely theoretical viewpoint, the  most evident
consequence of these oscillations is the persistent current flow
 $I (\Phi) = -c \partial F/\partial\Phi$ (here $F$ is the free energy) induced
by the magnetic flux that penetrates through quasi-1D conducting rings
\cite{Gunter}. In normal  metal-rings the oscillation period is  exactly the
quantum flux unit $\Phi_0 =  hc/e$, where $e$ is the electron charge, and
for fixed circumference $L$,
the oscillation amplitude can not  exceed  $I_0 \sim ev_F/L$. The
limiting value of the current, $I_0$, is determined by the  diamagnetic
electron
flow in a Fermi surface orbital and can be evaluated \cite{Kulik.BIL} from a
quite naive  model considering the free electron gas in a perfect ring
(i.e. a ring without any impurities; see also the review in
Ref.~\onlinecite{Imry}).

While the theoretical prediction of such persistent current flow in normal
metal-rings \cite{Buttiker}  was made some time ago,
 the corresponding experiments were performed only in the late 80's.
The measurement
of the persistent current was reported in Ref.~\onlinecite{Levy}, where the
induced magnetic moment in a stack of 10$^7$ copper-rings was
detected. The
period of oscillations was found to be $\Phi_0/2$, rather than $\Phi_0$. The
apparent puzzle was resolved by noting that the  contribution of the
fundamental
harmonic was averaged out over the whole ensemble of  rings in the system.
The effect of period halving can be easily understood in the free electron
model
assuming that the number of electrons, $N$, in a given ring rather then
chemical
potential is a conserved quantity. In this case the sign of the $n$th Fourier
harmonics entering the expression for the persistent current is  $(-1)^{Nn}$,
and after ensemble averaging (fluctuating $N$) all the odd terms are averaged
out. \cite{Cheung}

The Aharonov-Bohm oscillations of the magnetization with fundamental period
$\Phi_0$ were  first observed in a single
gold  ring. \cite{Chandrasekhar} The
amplitude of the oscillations  appeared to be unexpectedly large (on the
order of $I_0 \sim ev_F/L$) and
  this has triggered an avalanche  of theoretical papers
discussing the AB oscillations in isolated ``dirty" normal metal-rings.
The key
point here is that by properly taking into account the elastic scattering by
impurities,
 the dynamics of the charge carriers in a multichannel ring becomes
diffusive (inelastic processes are negligible at sufficiently low
temperatures). Therefore, a rather straightforward zero
temperature estimate suggests that $I_D\sim I_0\ell_D/L$, where $\ell_D$  is
the
electron mean-free path.
However,  the actual value of the current amplitude is
 greater than $I_D$ by several orders of magnitude in the  diffusive regime
where $\ell_D\ll L$, and this glaring disagreement has initiated a discussion
in
the literature.
\cite{Altshuler,Fabrizio94,Gogolin93,Kusmartsev94,Kato94,Vignale94,Yoshioka94,Smith92}

Metallic rings are intrinsically disordered, generally contain many
transverse channels and are in the diffusive regime.
In this paper we investigate the opposite limit of the effect of a single
impurity on a single-channel semiconductor
ring in the ballistic limit.  This limit
is instructive because it allows a reliable treatment of the
particle interactions.  Furthermore, recent progress
in semiconductor quantum wire technology has allowed the production of
clean low-channel number rings and measurement of their persistent
currents.\cite{Mailly}
Another important reason for studying this problem
is its connection to the problem of how the nature of the
 Coulomb blockade changes as the number of transverse channels in
a Luttinger liquid is varied.\cite{LIG:prl93}

The measurements of the magnetic moment in impurity-free rings
in Ref.~\onlinecite{Mailly} revealed
oscillations
with a fundamental period $\Phi_0$ and amplitude, $I_0$.
These results are in perfect agreement with  predictions
 based on the
free Fermi-gas model. At  first glance, this result is mysterious since at  low
charge density the Debye screening is weak and long range Coulomb forces  must
be a decisive factor for the quantum dynamics of the system. The effect of
electron-electron correlations on the persistent current of an isolated
metallic
ring were studied in  Ref.~\onlinecite{Ambegaokar} where it was
shown that in the
diffusive regime the interaction can enhance the  current. For the case of
ballistic rings, such a problem was addressed in Ref.~\onlinecite{Loss.prl}
building from the Luttinger model, and in Ref.~\onlinecite{Loss.prb} where the
QED radiative  corrections to  the current were calculated. The period and
zero-temperature amplitude of oscillations are  not affected by the Coulomb
interactions. This is as it should be according to general arguments.
\cite{Legget,Muller} Even the temperature dependence of the amplitude is in
fact the same as for the ideal gas. For an isolated ring with a fixed number of
electrons the influence of temperature is to destroy the oscillations. The
exponential suppression of the amplitude of the persistent current starts at a
crossover temperature,  $T_0 \sim \hbar v_F/ L$, which is of the order of the
energy-level spacing at  the Fermi energy for the free Fermi-gas model.

In this paper, we study the persistent current in a 1D ballistic ring in the
limit of strong Coulomb repulsion so that the (fluctuating)
 one-dimensional Wigner crystal (WC) representation of the Luttinger liquid
is valid.\cite{Likharev,Glazman.prb}
Note that the use of the name Wigner crystal here does not imply that
the system has long-range order. (It does not.)
Within a long wavelength approximation
the system
 can be viewed as an elastic chain of identical charges completely
described by three independent quantities:
the (electron) mass  $m$, the lattice
spacing $a$, and the sound velocity $s$. The magnitude of the  quantum
fluctuations are determined \cite{Glazman.prb} by a dimensionless parameter
$\alpha = \pi\hbar/msa$. In what follows, we take $\alpha \ll 1$, which
corresponds to a stiff crystal.

In the absence of crystal pinning (perfect ring), the persistent current
induced by the magnetic flux is caused by a uniform sliding of the Wigner
crystal as a whole. The elementary charge  transfer corresponds to a
translation of the crystal ring by the lattice constant $a$. Due to the fact
that
 the electrons at the different sites of the Wigner lattice are
indistinguishable, such a translation in a circular geometry is  equivalent to
carrying a single charge, $e$, along a closed loop. Therefore the period of the
corresponding Aharonov-Bohm oscillations is equal to $\Phi_0$, exactly as in
the
free Fermi-gas model.

Another simple argument tells us why the oscillation amplitude also
remains unchanged in the crystalline phase. The uniform translation of the
crystal by a lattice spacing $a$ results in a ``diamagnetic" current
$I_{WC}\sim
ev_0/L$, and the corresponding ``drift"  velocity $v_0 \sim \hbar/ma$ coincides
with the Fermi velocity for a gas of free spinless electrons at  the same
density. Clearly, this very circumstance explains why the cross-over
temperature  is of the same order for the two systems. These simple arguments,
of  course, can not  explain the exact coincidence of the zero-temperature
expressions in question. This fact follows from a general theorem.
\cite{Muller}

The Wigner crystal in a perfect ring does not get pinned and its persistent
currrent is (at $T = 0$) identical to the one carried by free electrons. In
order
to activate internal modes of the chain  one has to consider pinning effects.
In
the latter case, the charge  transport becomes of the tunneling type
(macroscopic quantum tunneling of the Wigner crystal ring) and the amplitude of
the persistent  current acquires a dependence upon both the pinning barrier
height and the  intrinsic  properties of the crystal.

The influence of the barrier on the quantum dynamics of the finite Wigner
chain depends on the strength of the pinning.
Defining $T_s \equiv \hbar s/L$  as the characteristic
energy scale in a chain of a finite length, $L$,
the strong pinning regime occurs for
$V_p \gg T_s/\alpha $.  In this regime the quantum fluctuations of
the displacement field are effectively cut off at the  intermediate scale
$\ell_0\sim a(T_s/\alpha V_p)\ll L$ (see Ref.~\onlinecite{Glazman.prb}) so
that the long-wavelength
 dynamics  ($k <\pi/\ell_0$) is quasiclassical and the functional
integral for the free energy can be evaluated by the method of ``steepest
descent".

Charge transfer along the ring can be viewed as a depinning of the
crystal by means of tunneling. This process has two distinct stages: (i)
``fast" tunneling of a small ring-segment   ($l_0 \ll L$)  through the
barrier,
(ii) ``slow" relaxation of the elastically deformed state developed in the near
barrier  region. It is the contribution of the relaxation stage that dominates
the tunneling action. \cite{Larkin} Hence, the magnitude of the persistent
current crucially
depends  on the elastic properties of the chain.
We will
show
that at low temperatures the expression  for  $I(\Phi)$   contains an
additional
small factor   $(T_s/\alpha V_p)^{1/\alpha}$ compared to the case of  a
perfect WC-ring. With increasing temperature the
effective tunnel barrier decreases and  the persistent current starts to grow.
Two competing mechanisms --- thermally activated tunneling and the temperature
destruction of quantum coherence due to the enhancement of destructive
interference --- lead  (for $\alpha\ll 1$) to the  appearance of a sharp
maximum in the temperature dependence of the  oscillation amplitude at   $T  =
T_m \simeq  0.5 T_s$.

Although at $T>T_s$ the destructive effect dominates and the magnitude
of the current is exponentially suppressed (this is also true for the ideal
crystal), it is partly compensated at $T\ll \alpha V_p$ by a growing
pre-exponential factor  $(T/\alpha V_p)^{1/\alpha}$.
This
$T^{1/\alpha}$ law  for tunneling of strongly correlated
electrons was predicted
in Refs.~\onlinecite{Glazman.prb} and \onlinecite{Kane.prb}.

A relatively weak suppression of the oscillations ensures that at temperatures
of the order of  $T\sim \alpha V_p$ the magnitude of the current
should coincide with its value for an ideal crystal.
Here the elastic
relaxation stage has no  effect since the initial length $\ell_0(T)$ of the
deformed segment is as large as the ring circumference $L$,  and the whole
transport is due to the uniform tunneling, $\exp(-V_p/T_s)$. The same
exponential factor determines the amplitude of the persistent current in a
perfect ring at   $T\sim \alpha V_p$.

With a further increase in temperature, the quantum depinning is replaced
by a  thermally activated depinning process, $V_p/T_s \to V_p/T$, and the
corresponding expression for $I(\Phi)$ acquires  at $T\gg \alpha V_p$  a
factor, $\exp(-V_p/T)$, that grows with temperature in addition to the
factor $\exp(-[\pi/2][T/\alpha T_s])$, which is a decreasing function of
temperature.

The above scenario relies on two assumptions: (i) the crystal is stiff,
$\alpha\ll 1$, and (ii) the
pinning is strong,  $V_p\gg T_s/\alpha$.
In the case of weak pinning the picture of macroscopic tunneling
of the Wigner crystal ring is changed drastically.
Firstly, the long-wavelength
 fluctuations of the displacement field are cut off now
but at the scale $L$  (in the weak pinning regime $\ell_0$ is larger than $L$).
The requirement of weak fluctuations puts  an upper bound on both the chain
length and the temperature:   $L\ll a{\rm e}^{1/\alpha}$, $T<T_s/\alpha$.
Secondly, for low temperatures, $T\ll T^*\simeq (\alpha T_s V_p)^{1/2}$, and
weak pinning the charge transport through the barrier goes entirely via
succesive tunneling rotations of the Wigner crystal ring as a whole by a
distance $a$.
 In  accordance with this, the zero-temperature amplitude of the
 current acquires (for $V_{p}\gg T_{0}$) a small tunneling factor
$\exp(-a\sqrt{2m(L/a)V_{p}}) \;=\; \exp(-\sqrt{2\pi V_{p}/T_{0}})\ll 1$.
Though at finite temperatures the tunneling rate is increased and the
competition between the thermally activated tunneling and the enhancement of
destructive interference leads to a maximum of the persistent current at
$ T\simeq T^* $, this  maximum is weakly pronounced. Loosely speaking,
the effect of moderate pinning, $\alpha T_s\ll V_p \alt T_s/\alpha$,
results in a suppression of the zero temperature amplitude of the
current and in a shift of the crossover temperature, $T_0$, to the
barrier dependent value $T^*\simeq \sqrt{\alpha T_{s}V_p}\;$. For very
weak pinning, $V_p\ll T_0$, the persistent current of a Wigner crystal ring
is indistinguishable
from that of non-interecting electrons.

 At high
temperatures, $T\gg {\rm max}\{\alpha T_s, (\alpha T_s V_p)^{1/2}\}$,
quantum depinning is replaced by a thermally activated depinning process, and
the conventional  picture of the Aharonov-Bohm oscillations in an ideal Wigner
crystal is recovered, up to a  slight correction due to the existence of the
barrier.

To summarize, the temperature dependence of the persistent current
differs drastically from the predictions based on the Fermi-gas model both in
the regime of strong and  moderately weak  pinning ($V_p > T_0$). To be more
specific, the temperature dependence exhibits a sharp maximum at $T\simeq 0.5
T_s$ in the case of strong pinning,  $V_p\gg T_s/\alpha$. For moderately weak
pinning, $T_0\alt V_p\alt T_s/\alpha $, the maximum is weakly pronounced, but
its
position is shifted to a potential-dependent value $T^*\simeq (T_0 V_p)^{1/2} >
T_0 $ which can be considered as the crossover temperature (instead of $T_0$)
in
this regime of pinning. In both cases the zero temperature amplitude is
strongly
suppressed.
 This conclusion is supported by the basic difference in the tunneling dynamics
of  strongly correlated and free electrons in an infinite one-dimensional
system. \cite{Kane.prl,Kane.prb}

For non-interacting electrons, the influence of a regular potential $V_p\ll
E_F$ --- where $E_F$ is the Fermi energy --- is negligible. \cite{Cheung}
Although in the presence of scatterers, not only  thermodynamic
quantities but also the energy-levels oscillate as a function of flux,
these oscillations generally come with alternating signs so that only the
energy
level closest to the  Fermi level in effect contributes to the current
oscillations. The latter is insensitive to the action  of a ``weak" ($V_p\ll
E_F$)  potential. The temperature dependence is also unaffected in this case.
\cite{Grincwajg}
Therefore, measurements of the persistent current in ballistic rings
may provide a deep insight into the quantum properties of strongly correlated
electron systems, and hence  offer a unique possibility of detecting and
 investigating properties
of a Luttinger liquid/Wigner  crystal in detail.

\section{Persistent Current in a Perfect Wigner Crystal Ring}

Let us consider a mesoscopic ring formed in a laterally confined
two-dimensional electron gas. For a low density of electrons ($n a_B\ll
1$, where $a_B$ is the Bohr radius)  weakly  screened Coulomb
interactions can lead to the formation of a 1D Wigner crystal.
\cite{Likharev,Glazman.prb,Jauregui} We will  study the response of
such a  strongly correlated electron system to an enclosed flux,
$\Phi$, of  magnetic field directed normal to the plane of the ring.

It is well known \cite{Byers} that in a non-simply connected space the free
energy, $F$, of the charged  particles acquires a flux-dependent part
resulting
in the appearence of an induced  magnetic moment. \cite{Bloch} This moment can
be related to the persistent current in the ring which is given by
\begin{equation}
I(\Phi) = -c \frac{\partial F(\Phi)}{\partial \Phi}			.
\label{one}
\end{equation}
For a non-superconducting ring the amplitude of the persistent current is a
decreasing function of the ring-circumference, $L$, and flux-induced effects
can be  detected only in samples of mesoscopic sizes.

Adopting the model proposed in Ref.~\onlinecite{Glazman.prb} we will describe a
1D Wigner crystal as an  elastic chain of spinless electrons. The input
parameters of the model are the crystal  period $a$, the electron mass, $m$,
and the sound  velocity $s$. In the  continuum limit the Lagrangian of such a
system in the presence of a pinning potential is \cite{Glazman.prb}
\begin{equation}
{\cal L} = \frac{ma}{8\pi^2} \left\{ \dot \varphi^2 - s^2(\varphi^\prime)^2
\right\} -V_p\delta(x) (1-cos \varphi)			 .
\label{two}
\end{equation}
Here $\varphi=2\pi u(x,t) /a$  is the dimensionless dynamical field of the
crystal  ($u (x,t)$ being  the local displacement of the crystal at the point
$x$  and
time $t$).
The notation $\dot \varphi$ and $\varphi^\prime$ indicate a
time- and spatial derivative, respectively. The  pinning potential (with
amplitude $V_p$) is assumed to be smooth on the scale of $a$ but well
localized
on the scale of the ring size $L$. Without loss of generality we place the
pinning  potential at point $x = 0$.

In low dimensional systems one should take into account quantum
fluctuations which can be very strong in the long-wavelength limit.
These fluctuations manifest themselves in  the logarithmic increase of
the correlator of the displacement field at long distances  ($x\gg a$)
and lead to the breakdown of crystalline order in infinite Wigner chains
even at zero  temperature. The magnitude of quantum fluctuations in a WC
can be characterized  \cite{Glazman.prb}
by a  dimensionless parameter $\alpha$
\begin{equation}
		\alpha = \frac{\pi \hbar}{msa}   .
\label{three}
\end{equation}
In what follows we will study a stiff (weakly-fluctuating) WC, which
implies that $\alpha\ll 1$.

In a perfect (or weakly pinned) WC, long-wavelength fluctuations are cut off at
a wavelength  on the order of the
crystal size $L$. It is physically
evident that we can asssume an ordered  crystalline structure as long as
the mean square fluctuations of the dimensionless field  $\varphi$,
\begin{equation}
		\langle \varphi^2 \rangle \sim \alpha
\int_{\pi/L}^{\pi/a}\frac{dk}{k}\coth\left(\frac{s\hbar}{2k_B T} k\right) ,
\label{four}
\end{equation}
 are small, so that $\langle \varphi^2 \rangle\ll 1$ ($T$ is the temperature).
For $T \to 0$ this inequality imposes a restriction in the form of an upper
bound on the chain length $L\ll a{\rm e}^{1/\alpha}$. One can easily check that
for such  samples the thermal
fluctuations are suppressed up to the temperature $T
\alt T_s/\alpha$   ($T_s \equiv \hbar s/L$). The situation is changed
drastically for a strongly pinned WC where an  ``intermediate" cut off scale
appears. \cite{Glazman.prb} The pinned WC will be studied in the next  section
and here we calculate persistent current in a perfect   ( $V_p = 0$ )   Wigner
crystal ring.

In a magnetic field, directed normal to the plane of the ring, the
one-dimensional
Lagrangian (\ref{two}) acquires an additional term,   ${\cal L}_{int}$,
which describes the Aharonov- Bohm interactions of the WC with the vector
potential of an electromagnetic field, $A_\varphi = \Phi/L$ ($\Phi$ is the
magetic flux through the ring). The AB interaction rewritten in terms of the
displacement field $\varphi$, takes the form of a total time derivative
\begin{equation}
{\cal L}_{int}  = \frac{\hbar}{L} \frac{\Phi}{\Phi_0}\dot{\varphi} ,
\label{five}
\end{equation}
and affects, as must be the case, only the quantum dynamics of the crystal.

The flux-induced persistent current,   Eq.~(\ref{one}),
is defined in terms of the
sensitivity of the  free energy of the ring to
magnetic flux.
For the following
analyses, it is convenient to  express the free energy, F,,
as a path integral
over quantum and thermal fluctuations of the  displacement field,
\begin{equation}
F = -k_B T \ln\left\{ \sum_{n=-\infty}^{\infty} (-1)^{n(N-1)} \int {\cal
D}\varphi_n {\rm e}^{-S_E[\varphi]/\hbar} \right\} .
\label{six}
\end{equation}
Here $S_E$ is the Euclidean (imaginary time) action for the Lagrangian
(\ref{two}), (\ref{five}), while $N$ is the  total number of electrons in
the chain. Since the arbitrary states  $\{\varphi(x,t)\}$ of the WC
which differ from each other by a constant
   $2\pi n$  ($n$ is an integer) are
physically  indistinguishable, one should impose ``twisted" boundary
conditions on the field $\varphi$ in imaginary time  (see e.g.
Ref.~\onlinecite{Krive})
\begin{equation}
\varphi_n (\tau + \beta, x) = \varphi_n (\tau, x) + 2\pi n .
\label{seven}
\end{equation}
This is why the path integration in Eq.~(\ref{six}) includes
an additional summation over  the homotopy
 index $n$ (winding
number), which classifies homotopically inequivalent
 trajectories.

The appearance in Eq.~(\ref{six}) of a sign-altering weight-factor, which
depends on the parity  of the total number, $N$, of electrons in the ring,
describes the ``parity effect" \cite{Legget,Loss.prl} in  mesoscopic
systems of strongly correlated electrons. This factor has a simple
physical  interpretation. One can view
the important ring-exchange
transformations of
the WC as a  homogeneous shift of length $a$   ($\Delta\varphi = 2\pi$)
supplemented by   ($N - 1$)   successive  permutations of pairs of
electrons. The corresponding extra phase  $\pi(N - 1)$ that appears  in
the many-particle wave function generates the sign-alternating factor in
Eq.~(\ref{six}). This  factor originates from the the
fact that the electrons in the WC-ring obey Fermi-statistics
and  is identical for free- 
 and strongly correlated electron systems. \cite{Legget,Loss.prl}
This factor does not appear in the infinite Luttinger liquid
problem\cite{Kane.prl,Kane.prb} but is essential to the physics of finite
rings.

For a perfect Wigner crystal ($V_p = 0$) the problem of calculating the
flux-dependent  part of the free energy, $\Delta F(\varphi)$, can be solved
exactly. The Lagrangian is ``quadratic" and  the extremal trajectories obeying
the boundary conditions (\ref{seven}) are $x$-independent linear  functions
of imaginary time  $\tau$,
\begin{equation}
\varphi_n (\tau,x)  = \frac{2\pi n}{\hbar\beta} \tau ,
\label{eight}
\end{equation}
($\beta \equiv 1/k_B T$). By substituting   Eq.~(\ref{eight}) into
Eqs.~(\ref{two}), (\ref{five}), and (\ref{six}) and then performing the
summation  over $n$, it is easy to find that
\begin{eqnarray}
		&&\Delta F(\Phi)  = -k_B T \ln \vartheta_3(\theta,q)
\label{nine} \\
&&q\equiv  \exp\left(\frac{\pi}{2}\frac{T}{T_0}\right), \quad
\theta = \Phi/\Phi_0 + \delta_N
\label{ten}
\end{eqnarray}
where  $\vartheta_3(\theta,q)$ is the Jacobi theta function, $T_0\equiv \hbar
v_F/L$   (in terms of WC parameters  $T_0 = \alpha T_s$, $T_s = \hbar s/L$),
and the parity
dependent term $\delta_N$ is 1/2 (0) for $N$ odd (even).

By making use of the asymptotic expressions for the $\vartheta_3$-function (see
e.g. Ref.~\onlinecite{Gradshtein}) we get  the desired formulae for the
persistent current at high- and low temperatures
\begin{equation}
I_{WC} \simeq \left\{ \begin{array}{ll} (-1)^N 2 I_0
\frac{T}{T_0}\exp\left(-\frac{\pi}{2}\frac{T}{T_0}\right)
\sin\left(2\pi\frac{\Phi}{\Phi_0}\right), &T\agt T_0 \\
I_0 \left(1 - 2\left\{\left\{\frac{\Phi}{\Phi_0} + \delta_{N}\right\}\right\}
\right) , & T\ll T_0.
\end{array}
\right.
\label{eleven}
\end{equation}
Here $\{\{x\}\}$ denotes the fractional part of $x$.  Thus the persistent
current carried by an ideal  Wigner crystal ring is a periodic function of flux
with period  $\Phi_0 = hc/e$ and amplitude $I_0 = e v_F / L$ at low
temperatures. The current has a paramagnetic character when there is  an even
number of elecrons in the ring (i.e. the induced magnetic moment is parallel to
the  external magnetic field) and diamagnetic for an odd number of electrons.
All these  properties of the persistent current exactly coincide with those
calculated using the model  of an ideal Fermi gas. \cite{Kulik.BIL}
The response of a perfect ring  (without impurity scattering) to
magnetic flux at $T = 0$ is essentially independent of the character  and
strength of Coulomb correlations among the electrons. For the Luttinger model
this was shown first in Ref.~\onlinecite{Loss.prl}.
A rigorous microscopic derivation of this fact
has been given by M\"uller-Groeling et al.\cite{Muller}

We call attention to the numerical factor in the exponent of
Eq.~(\ref{eleven}). According to this the temperature,
$T_{c}=(2/\pi)T_0=(2/\pi)(\hbar v_{F}/L)$,  that determines the
cross-over to the regime where the persistent current is suppressed by
loss of quantum coherence, $ I\propto \exp(-T/T_{c}) $, is twice as
large as for an ideal ring with Fermi-distributed electrons
\cite{Kulik.BIL,Cheung} (fixed chemical potential $\mu$). The appearence
of an additional factor of 2 in the cross-over temperature in our case
is connected with the statistical properties  \cite{Grincwajg,Kamenev93}
(thermal
averaging over a  canonical ensemble with a fixed number of particles) and is
weakly dependent on the electron-electron correlations. ~\cite{Loss.prl}

Now we need to consider the influence of fluctuations (temperature induced
``phonons" or plasmons in the WC) on the persistent current of a Wigner
crystal ring. If we have a rigid displacement of a ring by $ L/a = N$ steps
($N$
is the total number of electrons in the ring) the system returns exactly to its
original state, and so do the distortions. For $n\neq N$ we must move the
distortion to get the system back to its initial state $\varphi(x)\rightarrow
\varphi(x-na)$. This procedure implicitily takes into account the
finite period $a$ of the Wigner lattice and therefore goes beyond
the long-wavelength approximation used in Eqs.~(\ref{two}),(\ref{five}),and
(\ref{seven}), where all short wavelength processes have been included
in the sense that they have renormalized the bare parameters of the
effective Lagrangian (\ref{two}). Nevertheless, it seems interesting to
estimate to what extent our temperature corrections, Eq.(\ref{eleven}),
are valid.

To do this it is convenient to go over to momentum space, where the shift
$\varphi(x)\rightarrow\varphi(x+na)$ corresponds to a simple change of
phase in the Fourier transforms of the displacement field,
$\varphi_q\rightarrow
\exp(i\theta_q) \varphi_q,\; \theta_q \equiv qna $. The action in momentum
space reads
\begin{equation}
S_q = \frac{ma}{8\pi^2\hbar}\int_{-\hbar\beta/2}^{\hbar\beta/2}d\tau
\{|\dot{\varphi_q}|^2 + s^2 q^2|\varphi_q|^2\},
\label{G1}
\end{equation}
and  describes two uncoupled harmonic oscillators with frequency
$\omega_q =sq$. The extremal path corresponds to the classical
motion in an inverted potential subject to the boundary condition
$\;\varphi_q(\tau=-\hbar\beta/2)=\exp(i\theta_q)\varphi_q(\tau=\hbar\beta/2)\;$.
A straightforward calculation of the partition function for the model
action (\ref{G1}) gives
\begin{equation}
\frac{Z_q(\theta_q)}{Z_q(0)} = \left[ 1 + \frac{\sin^2(\theta_q/2)}
{\sinh^2(\beta\hbar\omega_q/2}\right]^{-1}.
\label{G2}
\end{equation}

If we introduce the thermal coherence length $\xi_T=\hbar s/T$, then according
to Eq.~(\ref{G2}) only modes with $q\xi_T \alt 1$ are significant and
we can use the small-$q$ approximation
\begin{equation}
\frac{Z_q(n)}{Z_q(0)} \simeq e^{-(na/\xi_T)^2},\;\; na \alt \xi_T .
\label{G3}
\end{equation}
Since the number of modes for which $q\xi_T \alt 1$ is of the order
of $L/\xi_T$, we readily get the following rough estimate of the
action for moving the distortion field by $n$ steps
\begin{equation}
S_d \simeq n^2 N\left(\frac{a}{\xi_T}\right)^3 \simeq n^2\frac{T}{T_0}
\alpha^3(\frac{T}{E_F})^2\;,
\label{G4}
\end{equation}
where $E_F$ is the Fermi energy. Comparing Eq.(\ref{G4}) with
Eqs.~(\ref{nine})-(\ref{eleven}), it is easy to see that the contribution of
phonons
to the persistent current of a thermally distorted Wigner crystal ring
is additionally suppressed by a factor $\alpha^3 (T/E_F)^2 \ll 1$
and can be safely neglected in what follows.
(It is an irrelevant perturbation.)

\section{Macroscopic Quantum Tunneling of a Wigner Crystal-Ring}

Now let us study the persistent current in a Wigner crystal ring
pinned
by a potential  barrier. A uniform sliding motion of the crystal is
impossible in this case and the charge  transport is inevitably
associated with macroscopic quantum tunneling.

The tunneling of strongly correlated electrons through a weak link in an
infinite one-
dimensional chain has been studied \cite{Kane.prl,Kane.prb} in the
framework of the Luttinger model.  It was shown that in the case of
repulsive interactions the electrons at zero temperature are  completely
reflected by the barrier even for an arbitrarily small height of the
barrier. The  authors of Ref.~\onlinecite{Glazman.prb}, in essence, came
to the same conclusion when studying the tunneling  of a WC through a
potential smooth on the scale $a$. One of the most surprising results
of this paper is the claim that the pinning of the
Wigner crystal suppresses the long-wavelength
 fluctuations of the displacement
field. Unlike the case of a perfect WC, the  fluctuations in the pinned
crystal chain get cut off at the intermediate scale $a \ll k_c^{-1} \ll
L$,  which can be found by minimizing the energy.
\cite{Glazman.prb} One finds
\begin{equation}
		k_c \simeq \frac{\pi}{a}\left( \frac{V_0}{ms^2}
\right)^{1/(1-\alpha)},
\label{twelve}
\end{equation}
where $V_0\ll ms^2$ is the bare amplitude of the pinning potential.
Therefore, even for an infinite  chain the potential --- regardless
of its initial strength --- is screened only partially
\cite{Glazman.prb}
\begin{equation}
		V_p \simeq V_0 \left( \frac{V_0}{ms^2}\right)^{\alpha/(1-\alpha)}
\label{thirteen}
\end{equation}
Since the tunneling action of the WC diverges logarithmically at long
distances, the  charge transport along the imperfect infinite 1D chain
vanishes at $T = 0$. For a crystal  of finite size the probability of
tunnelling is finite, thus providing a non-zero persistent  current in a
pinned WC-ring subject to a magnetic field.

In the following we will consider Eq.~(\ref{two}) to be an effective
Lagrangian. According to the  results of Ref.~\onlinecite{Glazman.prb}
short-wavelength fluctuations   ($k< k_c$) lead only to a
multiplicative  renormalization [Eq.~(\ref{thirteen})] of the amplitude
of the pinning potential. Therefore the renormalized
Lagrangian (\ref{two}) can serve as a starting point for studying the
long-wavelength quantum  dynamics of a WC-ring. In the limit $\alpha\ll
1$ (stiff Wigner crystal)  one can use quasiclassical approximations  in
calculating the path intergral, Eq.~(\ref{six}), for the free energy.

Tunneling of strongly correlated electrons --- described by the model
in Eq.(\ref{two}) ---
through a smooth barrier has been
considered thoroughly in Refs.~\onlinecite{Loss.prl} and
\onlinecite{Glazman.prb}. In the case
of strong pinning $\alpha V_p\gg T_s$ the process of macroscopic quantum
tunneling can be divided formally into two different stages: (i) a rapid
stage resulting in the formation of a local deformation in a small
region  $2\ell_0$ (see below) adjacent to the barrier, and (ii) a slow
process of elastic relaxation of the  deformed segment.

We shall first
calculate the
tunnel action and persistent current at zero temperature. At the rapid
stage of the
tunneling the displacement field, $\varphi$, ``hops" ($\Delta\varphi=
2\pi$) locally in the vicinity of the impurity. It is described by the
trial instanton-trajectory proposed in Ref.~\onlinecite{Larkin}. The
corresponding tunnel action, $A_h$,   is of the form
\begin{equation}
		A_h = \frac{\hbar}{\alpha}\left(C _1 + C _2 \frac{\alpha V_p}{T_h}\right),
\quad T_h \equiv \frac{\hbar s}{\ell_0}.
\label{fourteen}
\end{equation}
The exact values of the numerical constants $C _1 \sim C _2 \sim 1$ are not
essential
in what follows; $2 \ell_0$   is the initial length of the deformed segment of
the
WC. It can be found by minimizing the total action  (see below).

The second stage of tunneling describes the elastic relaxation of the
deformed state and the corresponding instanton solution obeys the free
equation of motion (in imaginary  time) for the displacement field
$\varphi$,
\begin{equation}
		\ddot\varphi + s^2 \varphi^{\prime\prime} = 0
\label{fifteen}
\end{equation}
The desired solution should satisfy the boundary conditions
\begin{eqnarray}
		&& \varphi_\pm (\tau= -\infty, x) -\varphi_\pm (\tau=\infty,x) = \pm 2\pi
\label{sixteen} \\
		&& \varphi_\pm (\tau, L/2) =  \varphi_\pm (\tau, -L/2)
\label{seventeen}
\end{eqnarray}
(the subscript, $ \pm $, labels trajectories with winding numbers $n = \pm 1$)
and represent the tunneling dynamics for $|x| \ge \ell_0$ and $|\tau| <
\tau_0 =\ell_0/s$; at the boundaries  it has to match the trial
function describing the first stage of the tunneling. A tunneling
trajectory which satisfies all these requirements has the form (Fig.~1a; see
also
Ref.~\onlinecite{Larkin})
\begin{equation}
\varphi_\pm = \pi \pm 2\arctan \left(\frac{(\tau-\tau_s)s}{|x|}\right)
\label{eighteen}
\end{equation}
Here $\tau_s$ is an arbitrary parameter which denotes the
location in time of
the instanton. The action for the relaxation stage (using the solution
(\ref{eighteen})) is
\begin{equation}
	A_r = \frac{\hbar}{\alpha}\ln\left( \frac{L}{2\ell_0} \right).
\label{nineteen}
\end{equation}

 It is easy to find that the minimum of the total action
$A_t = A_h + A_r $ corresponds to the following initial length, $2\ell_0$, of
the deformed segment
\begin{equation}
\ell_0 = \frac{\hbar s}{C _2\alpha V_p} \simeq a \frac{ms^2}{V_p} \gg a.
\label{twenty}
\end{equation}
For a stiff crystal this length  coincides with the lower bound
(\ref{twelve}) of the long wavelength  region, where the effective
Lagrangian (\ref{two}) is assumed to be valid. Thus our  calculation
can be  quantitatively reliable if $A_r\gg A_h$. This is the case of
strong pinning \cite{Larkin}  $\alpha V_p \gg T_s$, and
\begin{equation}
A_t =
\frac{\hbar}{\alpha}\left\{\ln \left(\frac{\alpha V_p}{T_s}\right)
+C _3 \right\}			 .
\label{twentyone}
\end{equation}
The numerical value of the constant $C _3$ in Eq.~(\ref{twentyone}) depends on
the first stage of the tunneling, which develops in the short-wavelength
region, and thus is beyond the accuracy of our  calculation.

With the help of Eq.~(\ref{twentyone}) it is easy to find the
flux-dependent part of the ground state  energy $E_0(\Phi)$. For a
stiff crystal ( $\alpha \ll 1$ )  the tunnel action is large,
$A_t/\hbar\gg 1$. Therefore we can use in our calculations the well-known
instanton dilute gas  approximation, where the tunneling trajectories are
``multi-step" solutions of Eq.~(\ref{fifteen}) (Fig.~1b; see e.g.
Ref.~\onlinecite{Rajaraman} as well as the similar calculation in
Ref.~\onlinecite{Bogachek.prb} of the Aharonov-Bohm  effect in
charge-density-wave conductors)
\begin{eqnarray}
E_0 &=& - k_B T \lim_{T\to 0}\left\{ \sum_{n=-\infty}^{\infty}
\sum_{n_+ ,
n_-=0}^{\infty}\frac{\left(K\sqrt{A_t/\hbar}\frac{T_s}{T}\right)^{n_+ +
n_-}}{n_+!n_-!} \times \right. \nonumber \\ &&\left.
{\rm e}^{-(A_t/\hbar)(n_+ + n_-) + i 2\pi\theta(n_+ - n_-)} \delta_{n, n_+ -
n_-}
\right\}.
\label{twentytwo}
\end{eqnarray}
Here $n_{+(-)}$  is the number instantons (anti-instantons), $K$ is
a numerical constant, $A_t$  is the tunnel action (\ref{twentyone}),
and  $\theta = \Phi/\Phi_0 + \delta_N $
(see Eq.~(\ref{ten}) ).

By doing the sums in Eq.~(\ref{twentytwo}) we readily get the well-known
general expression for the energy of a $\Theta$-vacuum in the dilute gas
approximation (see e.g. chapter 10 in Ref.~\onlinecite{Rajaraman})
\begin{equation}
		\Delta E_0(\Phi) = K T_s \sqrt{A_t/\hbar}{\rm e}^{-A_t/\hbar}
\cos \Theta .
\label{twentythree}
\end{equation}

In the case of strong pinning we only need to keep
the main (first)  term in the tunnel action (\ref{twentyone}).
Neglecting the inessential
prefactors one can derive from
Eq.~(\ref{twentythree}) the desired expression for the oscillating part
of the free energy at $T = 0$
\begin{equation}
\Delta F_{WC}(\Phi,T = 0) \sim (-1)^N T_s \left(\frac{T_s}{\alpha
V_p}\right)^{1/\alpha}{\rm
e}^{-C _3/\alpha}\cos\left(2\pi\frac{\Phi}{\Phi_0}\right). \label{twentyfour}
\end{equation}
We see that the amplitude of the persistent current in a pinned WC-ring
is much smaller  than in a perfect crystal. It decreases
rapidly with increasing
 crystal  size $\sim
(1/L)^{1+1/\alpha}$. Such a dependence for $\alpha \ll 1$ is
intermediate between the regimes of Aharonov-Bohm oscillations in
metals \cite{Imry} ($I \sim 1/L$)  and insulators
\cite{Kulik.88,Bogachek.JETP}   ($I\sim \exp (-L/ \ell_g$), $\ell_g
\equiv \hbar s/ \Delta_g$, $\Delta_g$ being the gap in the energy
spectrum at the Fermi level).

\section{Anomalous Temperature Dependence of a Persistent Current in a Strongly
Pinned  Wigner Crystal}

In a perfect Wigner crystal the temperature always acts to destroy the
Aharonov-Bohm oscillations (as in the case of a free Fermi-gas). Starting
from the cross-over temperature $T_0 \sim \hbar v_{F}/L$ the oscillation
amplitude is exponentially suppressed. In the presence of a pinning
potential the charge  transport is due to tunneling and the influence of
temperature can be more  complicated. From the argument that a finite
temperature facilitates tunneling, one  can expect an increase of the
persistent current in the low temperature regime.  Combining  this
effect with the loss of phase coherence at high temperatures due to the
enhancement of destructive interference, we may spectulate  that the
temperature dependence of the persistent current exhibits a maximum at
$T \sim T_s =\hbar s/L$. Below we will show that for a stiff crystal
this maximum develops into a sharp peak.

To prove this statement we have to calculate the flux-dependent
part of  the free energy of a WC-ring at non-zero temperatures. We will
again divide the process of depinning by tunneling into two stages: (i)
a rapid formation of a distorted region of size $2\ell_\beta \ll L$ in
the vicinity of the impurity, and  (ii) a slow relaxation of the elastic
tension during a ``time" $\tau_\beta = \hbar\beta$ ($\beta\equiv 1/T$).

In the previous section it was shown that in the case of strong
pinning,  $\alpha V_p\gg T_s$, the  dominating stage is the process of
elastic relaxation of the deformed state. Therefore we will not take
into account the temperature corrections to the action $A_h$, leaving
Eq.~(\ref{fourteen}) intact (now, however, the initial length of the
relaxing segment $2\ell_\beta (T)$ becomes temperature  dependent).

The main temperature dependence of the total tunneling action comes from
the relaxation  process due to the deformation of the Larkin-Lee
instanton, Eq.~(\ref{eighteen}). As before, we seek  the solution of the
free equation of motion, Eq.~(\ref{fifteen}), but with twisted boundary
conditions imposed on
the finite interval of imaginary time [$0, \beta$]
(see Eq.~(\ref{seven}) for $n=\pm 1$).  The desired solution should
transform into the Larkin-Lee instanton at low temperatures and
describe a uniform sliding of the WC-ring, Eq.~(\ref{eight}), in the
high temperature region.

It is easy to verify that the exact solution
is given by
\begin{equation}
\varphi_\sigma(x,\tau)=\pi +2\sigma\arctan\left[
     \coth\left(\frac{\pi|x|}{\hbar s\beta}\right)
\tan\left(\frac{\pi}{\hbar\beta}\left(\tau-\frac{\hbar\beta}{2}\right)
\right) \right],
\label{twentysix}
\end{equation}
where $\sigma = \pm 1$. This ``periodic" instanton solution holds when the
spatial coordinate, $x$, is outside the interval [$-\ell_\beta,
\ell_\beta$] containing the pinning potential, and for imaginary time
$|\tau - \hbar\beta/2|>\ell_\beta/s$. It should  be noted that our
solution, Eq.~(\ref{twentysix}) is similar to the so called ``colorons"
in quantum  chromodynamics \cite{Gross}, and it can be found using a
method proposed in Ref.~\onlinecite{Harrington}.

In the low- and high-temperature regions Eq.~(\ref{twentysix}) takes the
following
asymptotic  expressions
\begin{eqnarray}
\varphi_\pm &\simeq & \pi \pm 2\arctan \left(\frac{(\tau -
\tau_s)s}{|x|}\right),
\quad \beta\to\infty
\label{twentyseven} \\
\varphi_+ &\simeq &\frac{2\pi \tau}{\hbar\beta}, \quad \varphi_- \simeq 2\pi -
\frac{2\pi\tau}{\hbar\beta}, \quad \beta\to 0 .
\label{twentyeight}
\end{eqnarray}
(at low temperatures the interval of imaginary time is infinite,
$\beta\to\infty$, and  the center of the instanton can be placed at the
arbitrary point $\tau_s$). Equation (\ref{twentyseven}) is nothing but
the Larkin-Lee instanton. \cite{Larkin} The high temperature limit,
Eq.~(\ref{twentyeight}), describes a homogeneous instanton ($\varphi_+$)
and anti-instanton ($\varphi_-$) for the winding  numbers $n = \pm 1$.
Notice that at finite temperatures the center of the periodic instanton,
Eq.~(\ref{twentysix}), is a fixed quantity rather than an arbitrary
parameter (it is placed exactly in the middle of the imaginary time
interval) .

The action for the tunneling trajectory (\ref{twentysix}) is
\begin{equation}
A_p = \frac{\hbar}{\alpha}\left\{\ln\left[\frac{\sinh(\pi T/T_s)}{\sinh(2\pi
T/T_h)}\right] -
\frac{\pi}{2}\frac{T}{T_s}\left(1-2\frac{T_s}{T_h}\right)\right\},
 \label{twentynine}
\end{equation}
where $T_h \equiv \hbar s/\ell_\beta (T)$. The initial length,
$2\ell_\beta(T)$, of
the deformed segment which  minimizes the total action is equal to
\begin{equation}
\ell_\beta(T) = \frac{\hbar s}{2\pi T}\coth^{-1}\left(\frac{1}{2}
+\frac{C _2}{2\pi}\frac{\alpha V_p}{T}\right) .
\label{thirty}
\end{equation}
We will assume that $\ell_\beta \ll L$, a criterion which one can show
to be equivalent to a restriction on temperature, $T\ll V_p$.

By substituting Eq.~(\ref{thirty}) into Eqs.~(\ref{fourteen}) and
(\ref{twentynine})
one finds for the total tunneling action
\begin{eqnarray}
A_t(T) &=& \frac{\hbar}{\alpha}\left\{  C _1 +
\left( \case{1}{2} +\frac{C _2}{2\pi}\frac{\alpha V_p}{T} \right)
\coth^{-1}
\left(\case{1}{2} +\frac{C _2}{2\pi}\frac{\alpha V_p}{T} \right)
- \right. \nonumber \\
&& \left. - \frac{\pi}{2}\frac{T}{T_s}
+\ln\sinh\left( \pi\frac{T}{T_s} \right) +
\case{1}{2}\ln\left[ \left( \frac{C _2}{2\pi}\frac{\alpha V_p}{T}+\case{3}{2}
\right)
\left( \frac{C _2}{2\pi}\frac{\alpha V_p}{T} -\case{1}{2} \right)
\right]\right\}.
\label{thirtyone}
\end{eqnarray}
Here $C _1\sim C _2\sim 1$ (see Eq.~(\ref{fourteen}))  and
Eq.~(\ref{thirtyone}) is valid at temperatures $T\ll \alpha V_p$.  The
tunneling action, considered as a function of temperature, has a
minimum at $T = T_m \simeq 0.5  T_s$,  which is independent of the values
of the numerical constants
 $C _{1,2}$ ($T_m$ is a solution of the transcendental equation
$\coth(z) =1/2 +1/z$, where $z=\pi T/T_s$).
Thus it characterizes the relaxation stage of the macroscopic quantum
tunneling of a Wigner crystal ring. The extremal action,
 $A_{t}(T_m)$, divided by the
tunnel action at $T = 0$, Eq.~(\ref{twentyone}), is also entirely  determined
by the
relaxation stage. The minimum of the action corresponds to a maximum in
the persistent  current and hence
the temperature dependence of the
AB-oscillations in a strongly  pinned WC-ring turns out to be anomalous.

In the quasiclassical approximation (stiff WC: $\alpha\ll 1$, strong
pinning: $\alpha V_p \gg T_s$)
the trajectories with  smallest winding numbers ($n = \pm 1$)
give the  main contributions to the oscillating part of the free energy.
The persistent current in this case is determined by its fundamental
harmonic
\begin{equation}
I_{WC} (\Phi,T) = (-1)^N I_{WC}(T) \sin\left(2\pi \ \Phi/\Phi_0\right).
\label{thirtytwo}
\end{equation}
The amplitude of the current calculated for the ``periodic" instanton solution,
Eq.~(\ref{twentysix}), is
\begin{eqnarray}
I_{WC}(T) &=&
I_{WC}(0)\frac{T}{T_s}\exp\left[
 \frac{1}{\alpha}f\left(\frac{T}{T_s}\right)
                         \right], \quad
T_s\left(\frac{T_s}{\alpha V_p}\right)^{1/\alpha} \alt T \ll \alpha V_p
\nonumber \\
f(x) &=& \frac{\pi}{2} x - \ln \left[\frac{\sinh(\pi x)}{\pi x} \right] ,
\label{thirtythree}
\end{eqnarray}
where $I_{WC}(0)$ is the value at which the persistent current
saturates at zero temperatures (due to the finite length of the ring
providing an infrared cutoff scale $T_s$):
\begin{equation}
I_{WC}(0)
 = \frac{eT_{s}}{\hbar}\left(\frac{T_{s}}{\alpha V_{p}}\right)^{1/\alpha}
{\rm e}^{-C _3/\alpha}
\label{zerocurrent}.
\end{equation}

 According to
Eq.~(\ref{thirtythree})  the persistent current for $\alpha \ll 1$ peaks
(see Fig.~2) at $T = T_m\simeq 0.5 T_s$. The width of the peak is of the
order $\sqrt{T_{0}T_{s}}$. Even though the oscillation
amplitude  at the maximum is  much less than the value of the persistent
current in a perfect ring at the same temperature $ T_{m} $, the
temperature-induced suppression of the AB-oscillation is comparatively
weak since in the regime $T\gg T_s$
\begin{equation}
I_{WC}
\simeq \frac{eT}{\hbar}\left(
\frac{T}{\alpha V_p}\right)^{1/\alpha}
\exp\left(-\frac{\pi}{2}\frac{T}{\alpha T_s}\right) .
\label{thirtyfour}
\end{equation}
This is because of the thermal stimulation of tunneling, which gives
rise to the monotonically increasing
prefactor in Eq.~(\ref{thirtyfour}). The unusual
temperature dependence --- a power-law behaviour, $T^{1/\alpha}$ --- of
transport properties of strongly correlated  electrons was
discussed previously in Refs.~\onlinecite{Kane.prl} and
\onlinecite{Glazman.prb}.

Rigorously speaking, the expressions derived in this section are valid
at $T\ll\alpha V_p$.  Nevertheless, we can use them for a qualitative
analysis of the situation at $T\sim \alpha V_p$.  According to
Eq.~(\ref{thirty}) the initial length of the deformed segment in this
case is of the order  of the ring circumference   ($2\ell_\beta \to L$
at $T\to C _2\alpha V_p /\pi$) and the relaxation stage of the tunneling
vanishes. Now a shift of the WC-ring by tunneling a distance $a$
corresponds to a uniform motion of the crystal as a whole. We can
therefore expect that the persistent current will coincide with that
of a perfect ring. In fact, according to Eq.~(\ref{thirtyfour}), at $T
\sim \alpha V_p$
\begin{equation}
I_{WC}(T\sim\alpha V_p) \sim
\frac{\alpha e V_p}{\hbar} \exp\left(-{\rm
const.}\frac{V_p}{T_s}\right). \label{thirtyfive}
\end{equation}
This expression coincides (up to a numerical constant) with the
corresponding  result (\ref{eleven}) for the perfect WC-ring. With
a further increase of temperature, the  pinning can be removed by
thermal fluctuations and the AB-oscillations take place in a weak
pinning regime.

\section{Persistent Current in the Weak Pinning Regime}

So far we have studied the coherent properties of a Wigner crystal ring
assuming that --- despite the ``short-wavelength" ($k < k_c$)
screening of the bare potential $V_0$, --- the renormalized  height of
the potential barrier $V_p$, Eq.~(\ref{thirteen}), is large enough to
classify the tunneling dynamics of the WC-ring as being in the strong
pinning regime, where $V_p\gg T_s/\alpha \sim ms^2 a/L$. In this case the
coherent  properties of the WC are radically different from those in
a perfect ring up to  temperatures  $T \alt \alpha V_p$. Now we will
concentrate on the influence of
weak pinning ($V_p\ll T_s/\alpha$) on the magnitude of persistent current.

In the case of weak pinning, the influence of the barrier on the
intensity of quantum- and  thermal fluctuations of the
displacement field in a WC of a {\em finite}  length, $L$, is
negligibly small. Because of this we can invoke
the same restrictions
on the allowed  length of the rigid chain and the temperature interval
as in the case of the ideal crystal: $L\ll a {\rm e}^{1/\alpha}$. With
these limitations, we can neglect the renormalization of the height of
the potential barrier, $V_R = V_{p}\exp(-\frac{1}{2}<\varphi^2>) \simeq
V_{p} (L/a)^{\alpha} \simeq V_p $, and the Lagrangian (\ref{two})
can be regarded as a  quasiclassical model of a stiff ($\alpha\ll 1$)
Wigner crystal.

In the regime of weak pinning it is reasonable to use perturbation
theory for calculating the persistent current. The equation of motion in
 imaginary time for the dynamical field, $\varphi$,  takes the form
\begin{equation}
\ddot \varphi + s^2 \varphi^{\prime\prime} -
V_p\delta(x)\sin \varphi =0. \label{thirtysix}
\end{equation}
In zeroth order the tunneling trajectories are coordinate independent
functions $\varphi_n(\tau)$, which are linear in imaginary time,
Eq.~(\ref{eight}). Hence, the first order correction to the
Euclidean action of an ideal WC is
\begin{equation}
\Delta S_p^{(1)}/\hbar = V_p/T
\label{thirtyseven}
\end{equation}
This action describes the ordinary thermally activated charge
transport ``over the barrier".

In order to find the next order corrections we will look for a
perturbation solution of Eq.~(\ref{thirtysix}) of the  form
$\varphi_n(x,\tau) = \omega_n\tau +\psi_n(x,\tau)$, where
$\omega_n\equiv 2\pi n/\hbar\beta$ and the function
$\psi_n(x,\tau)$ satisfies periodic  boundary conditions
\begin{equation}
\psi_n(x,\tau+\beta) = \psi_n(x,\tau), \quad \psi_n(0,\tau)=\psi_n(L,\tau).
\label{thirtyeight}
\end{equation}
It is easy to find that in second order $\psi_n^{(2)}$ is
\begin{eqnarray}
\psi_n^{(2)}(x,\tau) &=& \frac{V_p}{2\omega_n s}
\frac{\sinh(\omega_n x/s) - \sinh(\omega_n(x-L)/s)}{\cosh(\omega_n L/s) -1}
\sin(\omega_n\tau)  \nonumber \\ &&
+\left(\frac{V_p}{4\omega_n s}\right)^2
\frac{\sinh(\omega_n L/s)\left[\sinh(2\omega_n L/s) -
\sinh(2\omega_n(x-L)/s)\right]}
{\left(\cosh(\omega_n L/s) -1\right)\left(\cosh(2\omega_n L/s) -1\right)}
\sin(2\omega_n\tau).
\label{thirtynine}
\end{eqnarray}
By substituting  $\varphi(x,\tau)$ into the Lagrangian (\ref{two})
and performing the spatial and (imaginary) time  integrals, one finds the
Euclidean action, $S_E$, of a weakly pinned WC  to be
\begin{equation}
S_E/\hbar = \frac{\pi}{2}\frac{T}{\alpha T_s} n^2 + n2\pi i \frac{\Phi}{\Phi_0}
+
\frac{V_p}{T} + \left(\frac{V_p}{T}\right)^2\frac{3\alpha}{4}\frac{\coth(\pi n
T/T_s)}{n}
\label{forty}
\end{equation}
(note that in the last term of Eq.~(\ref{forty})   $n \ne 0$).

The desired expression for the oscillating part of the free energy
is found by substituting the action (\ref{forty}) into
Eq.~(\ref{six}) and performing the summation over the homotopy
 index
$n$. It is easy to see from Eq.~(\ref{forty}) that at $T\to 0$ the
contributions of the terms stemming  from the pinning potential exceed
the action of the ideal ring and thus the perturbation  theory fails. In
the low temperature region  $T \ll (\alpha T_s V_p)^{1/2}$ the
calculation of the  persistent current again requires
non-perturbative
methods and will be carried out below.

We consider the most interesting case of moderately weak pinning, when
the barrier is still high enough that $\alpha T_{s} \alt V_{p} \ll
T_{s}/\alpha $. (For still weaker pinning the corrections to Eq.~(11) are
negligibly small.)
 It is easy to see that for (moderately) weak pinning
the tunneling of the WC-ring should be homogeneous. This is
because any distortion of the crystal structure results in an effective
barrier height $ V_{el} \agt ms^2 (a/L) \sim T_{s} /\alpha $, which
by far exceeds the applied potential $V_{p}$. Thus the trajectories
which describe tunneling in the weak pinning case are independent
of spatial coordinate and as a function of imaginary time have
 the form
illustrated in Fig.~1b. They consist of elementary tunnel rotations ($
\Delta\varphi =\pm 2\pi $), occurring at arbitrary points in imaginary
time and in arbitrary sequences subject to the only restriction that
the net rotation in the ``time" interval $[0,\beta]$ equals $2\pi$.

A single-instanton (anti-instanton) trial trajectory,
$\varphi_{+(-)}$, of the  desired form
\begin{equation}
\varphi_{+,-} = \pm 2\pi\frac{\tau}{\tau_{*}}\;,
\label{fortyone}
\end{equation}
where the parameter $\;\tau_{*}\ll\hbar/T\;$ defines the time scale of
the tunneling rotation (see below), can be justified as follows. It is
easy to verify that the one-parameter trajectory
\begin{equation}
\varphi(\tau,x;\tau_0) = 2\pi\frac{\tau}{\tau_0} + \tau_0\frac{\alpha V_p}
{\hbar}\sinh(\frac{2\pi}{\tau_0s}|x|)\sin(2\pi\frac{\tau}{\tau_0}),
\label{es}
\end{equation}
where $\tau_0$ is an arbitrary parameter, is an exact solution of the
nonlinear Eq.~(\ref{thirtysix}). The action for this set of trajectories
takes the form
\begin{equation}
A_{t}(\tau_0) = \frac{maL}{2\tau_0}+\tau_0 V_p +\frac{\alpha V_p^2}{8\hbar}
\tau_0^2 \sinh(2\pi\frac{L}{\tau_0 s}).
\label{Action}
\end{equation}

For weak pinning the last term in Eq.~(\ref{Action}) can be treated as
a perturbation and the extremal action is easily found by an iteration
procedure. After some strightforward algebra we get
\begin{equation}
\tau_{*} = \sqrt{\frac{maL}{2V_{R}}} ;
\label{tau}
\end{equation}

\begin{equation}
A_{t}(\tau_{*}) = \hbar \sqrt{2\pi\frac{V_{R}}{\alpha T_{s}}} \equiv a\sqrt
{2m\frac{L}{a}V_{R}} ,
\label{Ae}
\end{equation}
where the effective barrier height, $V_R$, is
\begin{equation}
V_R = V_p (1 + \frac{\pi}{4}\frac{\alpha V_p}{T_s}).
\label{VR}
\end{equation}
For weak pinning, $\alpha V_p\ll T_s$, the renormalization of the bare
potential, $V_p$, is small thus justifying the use of the trial instanton
trajectory (\ref{fortyone}) and the use of a perturbation
expansion in solving the equation for the minimum action. It
is worth noting that the last expression in Eq.~(\ref{Ae}) shows
explicitly that the process studied describes a homogeneous tunneling
of a Wigner crystal ring over a distance $a$.

Since in the low temperature region, where $T\ll T^{*}=(\alpha
V_{p}T_{s})^{1/2}$, the characteristic time, $\tau_{*}$, of an
elementary tunneling rotation, $ \Delta\varphi =\pm 2\pi $, satisfies the
inequality $ T\tau_{*} \ll \hbar $, one can regard the successive
rotations as uncorrelated events and may use an instanton dilute gas
approximation when deriving the flux-dependent part of the free energy.
Thus the instanton described by Eqs.~(\ref{fortyone}) and (\ref{tau})
plays the same role in the weak pinning regime as the Larkin-Lee
instanton, Eqs.~(\ref{eighteen}) and (\ref{twentyone}), does in the
strong pinning regime. From Eqs. (23) and (24) we immediately get the
zero-temperature limit of the persistent current ($\alpha T_{s}\alt
V_{p}\ll T_{s}/\alpha $) as
\begin{equation} I_{wp}(T\rightarrow 0)\sim
(-1)^N I_{0}\left(\frac{V_{R}}{\alpha T_{s}}
\right)^{1/4}\exp\left(-\sqrt{2\pi\frac{V_{R}}{\alpha T_{s}}}\right)
\sin\left(2\pi\frac{\Phi}{\Phi_{0}}\right) .
 \label{iwp}
\end{equation}
Notice the exponential supression of the Wigner crystal current by the
impurity potential. This is in sharp contrast to the case of
noninteracting electrons, where  even a strong potential (of the order of
the Fermi energy) only leads to a power-law supression. \cite{Cheung}

The qualitative picture described above enables one to speculate about
the temperature dependence of the persistent current in the weak pinning
regime. Since temperature helps tunneling, we may expect
an anomalous temperature dependence of the persistent current of a weakly
pinned Wigner crystal ring.

It is physically evident that at finite temperatures the effective barrier
for tunneling is smaller than
at $T=0$. One can easily take this effect
into account by making use of a one-particle analogy of Wigner
crystal tunneling. In our case (see Eq.~(\ref{Ae})) the stiff crystal with
a total mass $mN$ will tunnel through the barrier at $T\neq 0$ via
thermally excited states. The corresponding temperature dependent action,
$A_{T}$, for $T\ll T^{*}$ can be estimated as follows (see e.g.
Ref.~\onlinecite {Glazman.prb})
\begin{equation}
e^{-A_{T}/\hbar} \simeq \int_0^{\varepsilon^{*}}\frac{d\varepsilon}{T}
\exp\left(-\frac{\varepsilon}{T}-\sqrt{\frac{2\pi}{\alpha T_{s}}(V_{p}-
\varepsilon )}\right)
\label{AT}
\end{equation}
(here we neglect the distinction between $V_R$ and $V_p$).

At low temperatures the upper limit of integration in Eq.~(\ref{AT}) is
irrelevant and we readily get the finite temperature corrections to
Eq.~(\ref{Ae}) as
\begin{equation}
A_{T}\simeq \hbar\sqrt{2\pi\frac{V_{p}}{\alpha T_{s}}}\left(1-\frac{1}{2}
\frac{T}{V_{p}}\right)\;\; ,\; \; T\ll V_{p}.
\label{ts}
\end{equation}
The temperature corrections to the tunneling action --- which are linear
in $T$ at low temperatures, much as in the case of strong pinning,
Eq.~(30) --- diminish the action and thus lead to an
enhancement of the persistent current.
The increase of persistent current amplitude saturates at a temperature
$T\simeq T^* =\sqrt {\frac{2}{\pi}V_p T_0}$ when the competing
temperature-induced loss of quantum coherence becomes important.
In contrast to the case of strong pinning the difference between the
tunneling actions at the mimimum and at zero temperature is not large
($\Delta A_t/\hbar \sim O(1)$). The maximum in the amplitude
of the persistent current is therefore weakly pronounced. Thus the only
distinctive feature of the temperature dependence of a moderately pinned
($ V_p \gg T_0 $) Wigner crystal ring --- as compared to the ideal Fermi
gas --- is a shift of the crossover temperature $ T_0\rightarrow T^*
\gg T_0$. For very weak pinning, $V_p\ll T_0$, even this distinction
vanishes and the response of a Wigner crystal ring to magnetic flux
approaches that of non-interacting electrons.

 The high-temperature region,
$T \agt {\rm max}\{\alpha T_s,\sqrt{\alpha T_s V_p}\}$,
can be described by perturbation theory, Eq.~(\ref{forty}). In this
case the main contribution to the free energy comes
 from the
trajectories with smallest winding  numbers, $n = \pm 1$. Hence,
according to Eqs.~(\ref{forty}) and (\ref{six}), one finds
\begin{equation}
\Delta F(\Phi) \simeq 2T(-1)^N\exp\left\{
-\frac{\pi}{2}\frac{T}{\alpha T_s} - \frac{V_p}{T} -\frac{3\alpha}{4}
\coth\left(\pi\frac{T}{T_s}\right)\left(\frac{V_p}{T}\right)^2\right\}
\cos\left( 2\pi\frac{\Phi}{\Phi_0}\right).
\label{fortytwo}
\end{equation}
We see that at high temperatures weak pinning leads to additional suppression
of the current in a perfect ring, Eq.~(\ref{eleven}),
and this potential-dependent
contribution decreases with increasing temperature.

 We gratefully acknowledge discussions with L. Glazman, A. Nersesyan,
A. Sj\"{o}lander, and A. Zagoskin. This work was supported by the Swedish
Royal Academy of Sciences, the Swedish Naural Science Research Council,
by the NSF through grant DMR-9113911, and by Grant U2K000 from the
International Science Foundation. S.G and I.K. acknowledge the
hospitality of the Department of Applied Physics, CTH/GU.

\begin{figure}
\caption {Sketch of tunneling trajectories of the dynamical displacement
field $\varphi_n(\tau)$ of the Wigner crystal-ring. The trajectories in
(a) correspond to `single-step' tunneling and winding number $n=1$ (full
curve) and $n=-1$ (dashed curve). In (b) an example of a $n=1$ `multiple-step'
tunneling trajectory is given.
}
\end{figure}

\begin{figure}
\caption{Temperature dependence of the normalized persistent current in a
strongly pinned Wigner crystal of different stiffness (measured by
$\alpha^{-1}=2msa/h$; $T_s=\hbar s/k_BL$, see text). The sharp peak for stiff
crystals is a result of a
competition between two effects of temperature: a reduced renormalized
tunneling barrier and a loss of quantum coherence.
}
\end{figure}

\end{document}